\documentclass[twocolumn,showpacs,preprintnumbers,amsmath,amssymb]{revtex4-1}

\usepackage{hyperref}
\usepackage{graphicx}
\usepackage{graphics}
\usepackage{amsmath}

\begin{document}
\title{ Phase shift in an atom interferometer induced by the additional laser lines of a Raman laser generated by modulation}

\author{Olivier Carraz}
\author{Ren\'{e}e Charri\`{e}re}
\author{Malo Cadoret}
\altaffiliation{Present address: Laboratoire Commun de M\'{e}trologie LNE-CNAM,
61 rue du Landy, 93210 La plaine Saint Denis, France}
\author{Nassim Zahzam}
\author{Yannick Bidel}
\email{yannick.bidel@onera.fr}
\author{Alexandre Bresson}
\affiliation{ONERA, BP 80100, 91123 Palaiseau Cedex, France}

\begin{abstract}
The use of Raman laser generated by modulation for light-pulse atom
interferometer allows to have a laser system more compact and robust. However,
the additional laser frequencies generated can perturb the atom interferometer.
In this article, we present a precise calculation of the phase shift induced by
the additional laser frequencies. The model is validated by comparison with
experimental measurements on an atom gravimeter. The uncertainty of the phase
shift determination limits the accuracy of our compact gravimeter at
$8\times10^{-8}\,\text{m}/\text{s}^{2}$. We show that it is possible to reduce
considerably this inaccuracy with a better control of experimental parameters
or with particular interferometer configurations.
\end{abstract}

\maketitle

\section{Introduction}

Light-pulse atom interferometry \cite{light pulse inter} is a promising
technology to obtain highly sensitive and accurate inertial sensors. Laboratory
experiments have already demonstrated state of the art performances for
gravimeter \cite{gravi peter, gravi syrte}, gradiometer \cite{gradio} and
gyroscope \cite{gyro, gyro syrte}. Inertial sensors start to be tested in
mobile platforms such as a plane \cite{ice nature} or a truck \cite{gradio
camion}. An important research effort has still to be done in order to have
instruments and particularly the laser system more compact and robust for
practical applications like navigation \cite{navigation}, space mission
\cite{space, SAGAS, MWXG, STE QUEST}, gravity mapping and monitoring
\cite{etna}, subsurface detection \cite{subsurf det}.

The key point of light-pulse atom interferometer is the stimulated Raman
transition between two stable states of the atom \cite{stimulated raman
transition}. This transition is driven with two counter-propagating lasers with
a frequency difference corresponding to the energy difference between the
stable states. This transition allows to make the equivalent of beam splitters
and mirrors for matter waves. To generate this Raman laser, two phase locked
lasers are generally used \cite{laser syrte}. Another solution consists in a
laser which is intensity or phase modulated at the frequency difference between
the ground states. This technology gives a laser system more compact (only one
laser is needed) and more robust (no phase lock). Moreover, the frequency noise
of the laser is not reported into the Raman phase noise \cite{limit sensitivity
grav}. The drawback of this technique is the presence of additional laser
frequencies which can perturb the atom interferometer. The realization of an
atom interferometer using a modulated Raman laser has been demonstrated
experimentally \cite{laser onera, ice nature} but with an additional phase
shift which have not been quantified yet.

In this article, the perturbations induced on an atom interferometer by the
additional frequencies present in a modulated Raman laser are studied in
detail. In the first part, we show a precise calculation of a stimulated Raman
transition with a modulated Raman laser. In the second part, we calculate for a
Mach-Zehnder interferometer the phase shift induced by the additional laser
frequencies. In a third part, we compare the results of our calculation to
experimental measurements of the phase shift. Finally, we present the
limitation in accuracy of an atom interferometer using a modulated Raman laser.
We also show a configuration where the inaccuracy induced by the additional
laser frequencies is considerably reduced.

\section{Stimulated Raman transition with a modulated laser}

\begin{figure}[h]
 \includegraphics[scale=0.65]{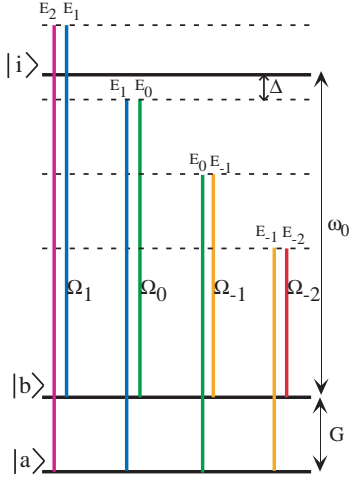}
 \caption{\label{spectrum}Atomic level system and Raman transitions.}
\end{figure}

\begin{figure}[h]
 \includegraphics[scale=0.65]{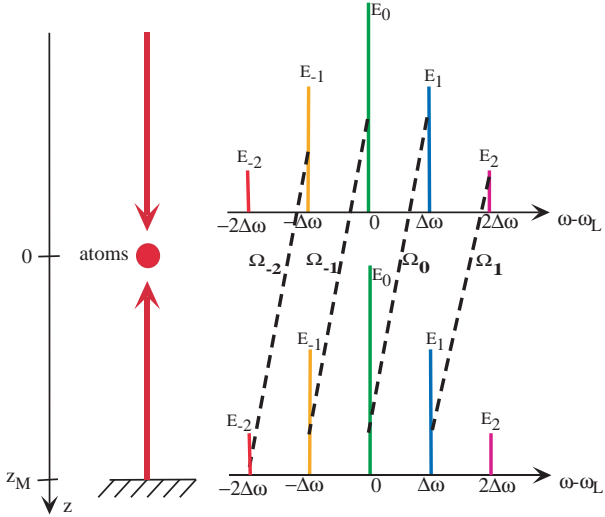}
 \caption{\label{configraman}Raman transition with a retro-reflected modulated laser. The dot lines correspond to the couple of laser lines driving the
 counter-propagating Raman transition considered in this article.}
\end{figure}

The atom corresponds to a $\Lambda$-type three-level system with two lower
levels $|a\rangle$ and $|b\rangle$ separated by an energy $\hbar
 G$ and an excited state $|i\rangle$ separated from the level $|b\rangle$ by $\hbar \omega_0$
 (see Fig. \ref{spectrum}). The atom interacts with a laser retro-reflected by a
mirror at the position $z_M$. The spectrum of the laser is composed of lines
separated in frequency by $\Delta\omega$. This kind of spectrum is obtained
with a laser modulated in amplitude or in phase with a frequency
$\Delta\omega$. The electric field seen by the atom can be written as:
\begin{eqnarray}\label{Electric field}
E=\sum_{n=-\infty}^\infty E_n \cos\left((\omega_L+n\Delta\omega)(t-z/c)+\varphi_n\right)\nonumber \\
+E_n\cos\left((\omega_L+n\Delta\omega)(t+(z-2z_M)/c)+\varphi_n\right)
\end{eqnarray}
When the frequency of modulation $\Delta\omega$ is close to the frequency
difference of the two lower levels $G$, each couple of laser lines separated by
$\Delta\omega\simeq G$ drives stimulated Raman transitions between the two
lower states (see Fig. \ref{spectrum}). In this article, one considers only
two-photon Raman transition with counter-propagating beams with the higher
frequency beam propagating downward (see Fig. \ref{configraman}). If the
velocity of the atoms is big enough, the other kinds of Raman transition
(co-propagating and counter-propagating with opposite direction) are out of
resonance thanks to the Doppler effect and can be neglected. The couple of
counter-propagating beams with frequencies $\omega_L+(n+1)G$ and $\omega_L+nG$
is coupling the state $|a,p\rangle$ to the state $|b,p+\hbar k_{\mathrm{eff}}+n
\hbar\Delta k\rangle$ where $k_{\mathrm{eff}}=(2\omega_L+\Delta\omega)/c$ and
$\Delta k=2\Delta\omega/c$. The effective Rabi frequency $\Omega_n$ associated
with this transition is equal to :
\begin{equation}
\Omega_n=\frac{\Omega^*_{(n+1)ai}\Omega_{nbi}}{2(\Delta+n\Delta\omega)}=\frac {E_{n+1}E_n}{E_1 E_0}\frac{\Delta}{\Delta+n\Delta\omega}\,\Omega_0
\end{equation}
where $\Delta=\omega_L-\omega_0$ and $\hbar \Omega_{nxi}=-\langle i|d_{xi}
E_n|x\rangle$ with $x=a\,\text{or}\,b$.

\begin{figure}[h]
 \includegraphics[scale=0.55]{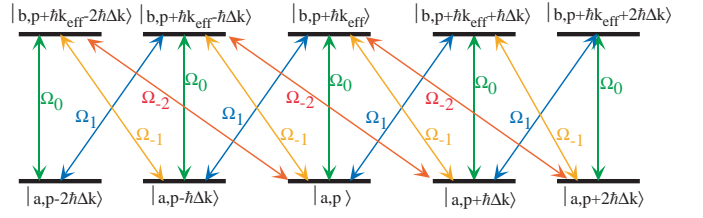}
 \caption{\label{family}Family of quantum states coupled with a modulated Raman laser.}
\end{figure}

The family of quantum states coupled with these Raman transitions is therefore
$|a,n\rangle=|a,p+n\hbar\Delta k\rangle$ and $|b,n\rangle=|b,p+\hbar
k_{\mathrm{eff}}+n\hbar\Delta k\rangle$ (see Fig. \ref{family}). Compared to
the case of a Raman laser with only two laser lines where two quantum states
are coupled, a modulated Raman laser couples an infinite number of quantum
states. A Raman transition with a modulated laser is therefore not equivalent
to a simple beam splitter but can be seen as a multiple beam splitter.

The calculation of the transition matrix will be done within the following
approximations. The kinetic energy difference induced by the term $\Delta k$
will be neglected. In typical atom interferometer ($G/2\pi\lesssim
10\,\text{GHz}$, $v\lesssim 1\,\text{m/s}$), the Doppler effect associated with
$\Delta k$ is below $67\,\text{Hz}$ and can be neglected for typical
interaction time of $10\,\mu\text{s}$. We will therefore consider that the
states $|a,n\rangle$ together with the states $|b,n\rangle$ are degenerated in
energy. We also suppose that the Raman laser is at resonance i.e. the frequency
of modulation $\Delta\omega$ is equal to the frequency difference between the
states $|a,n\rangle$ and the states $|b,n\rangle$ eventually affected by the
light shift. Within this approximation and by adiabatically eliminating the
excited state and using the rotating wave approximation, the system can be
described by the following effective Hamiltonian :
\begin{eqnarray}
H=\frac{\hbar}{2} \sum_n\sum_m  \Omega_m e^{i\phi_m} |b,n+m\rangle\langle a,n|\nonumber \\+ \Omega_m e^{-i\phi_m} |a,n-m\rangle\langle b,n|
\end{eqnarray}
where $\phi_m$ is the phase of the Raman laser associated with the couple $m$
of Raman laser :
\begin{equation}
\phi_m=\varphi_{m}-\varphi_{m+1}-(k_{\mathrm{eff}}+m\Delta k)z_M+\frac{\Delta k}{2} z_M
\end{equation}
If a free fall frame is used, one should replace in this expression $z_M$ by
$z_M-\frac{1}{2}g\,t^2$. To compensate the Doppler shift induced by gravity,
one generally applies a frequency chirp $\alpha$ to the Raman frequency. In
this case, one should add to the phase $\phi_m$ the term $-\alpha\, t^2/2$. The
terms $\frac{\Delta k z_M}{2}$ and $k_{\mathrm{eff}}z_M$ are constant and can
be ignored in the following. If the intensity or phase modulated laser
propagates through a dispersive medium, $\varphi_{m+1}-\varphi_{m}$ is not null
and is proportional to $m$. This term is equivalent to a modification of the
mirror distance $z_M$ and can be ignored. Finally, the expression of $\phi_m$
can be written into a term independent of $m$ and a term proportional to $m$ :
\begin{equation}
\phi_m=A+mB
\end{equation}
with
\begin{eqnarray}
A=\frac{1}{2}\left(k_{\mathrm{eff}}g-\alpha\right)\,t^2\nonumber\\
B=\Delta k \left(\frac{1}{2}g\,t^2-z_M\right)
\end{eqnarray}

The evolution operator of the atom interacting with the Raman laser is given
by:
\begin{equation}
U=e^{-i\frac{H t}{\hbar}}
\end{equation}
One can show that the elements of the evolution operator can be written as :
\begin{eqnarray}\label{transition coeff}
\langle a,n+m|U|a,n\rangle&=&t_m\,e^{i\,m\,B}\nonumber\\
\langle b,n+m|U|b,n\rangle&=&t_{-m}\,e^{i\,m\,B}\nonumber\\
\langle b,n+m|U|a,n\rangle&=&-i\,r_m\,e^{i\,A}\,e^{i\,m\,B}\nonumber\\
\langle a,n+m|U|b,n\rangle&=&-i\,r_{-m}\,e^{-i\,A}\,e^{i\,m\,B}
\end{eqnarray}
The coefficient $r_m$ and $t_m$ are real and can be calculated numerically by
truncating the number of states coupled and the number of laser lines.
Typically, we perform the calculation with 22 quantum states and 5 laser lines.
The calculation of the transition amplitude is done for the interaction times:
$\tau=\frac{\pi}{2\Omega_0}$ and $\tau=\frac{\pi}{\Omega_0}$ corresponding to
$\pi/2$ and $\pi$ pulses without additional laser lines (beam splitter and
mirror). The result of the calculation is presented in the table
\ref{transition} for the case of stimulated Raman transition on Rubidium 87.

We can notice that the probability transition into parasite states ($n\neq
0$)is small but not negligible. We should therefore take into account the
additional paths engendered by these parasite transitions in order to determine
precisely the phase of an atom interferometer.

\begin{table}[h]
\begin{tabular}{|c|c|c|c|c|}
\hline
 \rule{0cm}{0.4cm}n&$t^{\frac{\pi}{2}}_n$&$r^{\frac{\pi}{2}}_n$&$t^{\pi}_n$& $r^{\pi}_n$\\
\hline
 \rule{0cm}{0.5cm}-2&$\sqrt{1.1\times10^{-5}}$&$-\sqrt{8.3\times10^{-5}}$&$\sqrt{1.3\times10^{-4}}$&$-\sqrt{1.5\times10^{-4}}$ \\
\hline
\rule{0cm}{0.5cm}-1&$\sqrt{1.2\times10^{-3}}$&$-\sqrt{3.2\times10^{-3}}$&$\sqrt{9.2\times10^{-3}}$&$-\sqrt{1.1\times10^{-3}}$  \\
\hline
\rule{0cm}{0.5cm}0&$\sqrt{0.497}$&$\sqrt{0.497}$&$-\sqrt{3.0\times10^{-6}}$&$\sqrt{0.979}$\\
\hline
\rule{0cm}{0.5cm}1&$\sqrt{1.2\times10^{-3}}$&$-\sqrt{1.5\times10^{-4}}$&$\sqrt{9.2\times10^{-3}}$&$\sqrt{9.1\times10^{-4}}$  \\
\hline
\rule{0cm}{0.5cm}2&$\sqrt{1.1\times10^{-5}}$&$-\sqrt{1.8\times10^{-8}}$&$\sqrt{1.3\times10^{-4}}$&$\sqrt{1.3\times10^{-5}}$ \\
\hline
\end{tabular}
\caption{\label{transition}Transition amplitude with a phase modulated Raman
laser with $G/2\pi=6.8\,GHz$, $\Delta/2\pi=0.7\,GHz$, $E_n=J_n(1.25)$ where
$J_n$ is the Bessel function of the first kind of order $n$.}
\end{table}

\section{Phase shift in a Mach Zehnder interferometer induced by the additional laser lines of a modulated Raman laser}

In this section, we present the calculation of a Chu-Bord\'{e} interferometer
using a modulated laser for stimulated Raman transition. This interferometer
consists of three Raman laser pulses of duration $\tau$, $2\tau$ and $\tau$
($\Omega_0 \tau =\pi /2$) separated in time by T. This interferometer is
equivalent to an optical Mach Zehnder interferometer where the first and last
pulses act as beam splitters and the second pulse acts as a mirror. The
evolution operator of this interferometer is given by:
\begin{equation}\label{evol_interf}
U=U_3\cdot U_L \cdot U_2
\cdot U_L \cdot U_1
\end{equation}
where $U_n$ is the evolution matrix for the $n^{th}$ laser pulse which has been
calculated in the previous section and $U_L$ is the free evolution during the
time T. The free falling frame will be used for the calculation. The
gravitational potential is therefore not included in the free evolution and one
has only the internal energy and the kinetic energy :
\begin{equation}
U_L=\sum_n e^{-i\omega_{an}T}|a,n\rangle \langle a,n|+e^{-i\omega_{bn}T}|b,n\rangle \langle b,n|
\end{equation}
with:
\begin{eqnarray}\label{omega free}
\omega_{an}&=&\frac{(p+n\hbar\Delta k)^2}{2\hbar M}\nonumber\\
\omega_{bn}&=&\frac{(p+n\hbar\Delta k+\hbar k_{eff})^2}{2\hbar M}+G
\end{eqnarray}
where $M$ is the mass of the atom.

We assume that the atom is initially in the internal state $|a\rangle$ and has
a momentum probability amplitude $\varphi(p)$ :
\begin{equation}\label{initial_state}
|\Psi_0\rangle = \int \varphi(p) |a,p\rangle dp
\end{equation}
The momentum probability amplitude $\varphi(p)$ will be modeled by a gaussian
wave packet with a mean momentum $p_0$, a width given by the temperature of the
atoms $T_a$ and centered at the position $z=0$.
\begin{equation}
\varphi (p)=\frac{1}{(2\pi\, M\, k_B T_a)^{1/4}}\, \exp{\left(-\frac{(p-p_0)^2}{4\, M\, k_B T_a}\right)}
\end{equation}
 After the interferometer sequence, the probability to be in the state
$|a\rangle$ is given by :
\begin{equation}
P_a=\int |\langle a,p|U|\Psi_0\rangle|^2dp
\end{equation}
By inserting equation (\ref{initial_state}) into the expression of $P_a$ and
using the family of quantum states coupled by U (see Fig. \ref{family}), one
obtains:
\begin{equation}
P_a=\int \left|\sum_n \varphi(p+n\hbar \Delta k) \langle a,p|U|a,p+n\hbar\Delta k\rangle\right|^2 dp
\end{equation}
We assume now that the atomic coherence length $l_c=h/\sqrt{M\,k_B\,T_a}$ is
small compared to the microwave wavelength $2\pi/\Delta k$. This approximation
is completely valid for atoms at a temperature of $\sim 1 \mu \text{K}$ ($l_c
\sim 0.5\,\mu \text{m}$) and with $G/2\pi\lesssim 10\,\text{GHz}$ ($2\pi/\Delta
k\gtrsim 1.5\,\text{cm}$). Within this approximation, one can write
$\varphi(p+n\hbar\Delta k)\simeq \varphi(p)$ and one obtains :
\begin{equation}
P_a=\int{ \left|\sum_{n} \langle a,p |U|a,p+n\hbar \Delta k\rangle  \right|^2 |\varphi(p)|^2 dp}
\end{equation}
By decomposing U in free evolution and laser interaction (Eq.
(\ref{evol_interf})), one obtains :
\begin{equation}
P_a=\int |\varphi (p)|^2 |C_{ab}+C_{ba}+C_{aa}+C_{bb}|^2dp
\end{equation}
with:
\begin{eqnarray}\label{Cij}
C_{ij}=\sum_{n'',n',n}
 e^{-i\,(\omega_{jn''}+\omega_{in'})T}\,
\langle a,0|U_3|j,n''\rangle
\langle j,n''|U_2|i,n'\rangle \nonumber\\
\langle i,n'|U_1|a,n\rangle\nonumber\\
\end{eqnarray}
In this expression, $C_{ij}$ represents the probability amplitude of an atom to
follow the path $a\rightarrow i \rightarrow j \rightarrow a$ where i,j = a or
b. We will consider only the interference between the term $C_{ab}$ and
$C_{ba}$.  The interferences with the terms $C_{aa}$ and $C_{bb}$ vanish if the
path separation $D=\hbar k_{eff}T/M$ is much bigger than the coherence length
of the atoms. This assumption is perfectly verified in practical situation with
$T>1\,\text{ms}$ ($D>100\,\mu\text{m}$) and atoms at a temperature $T_a \sim 1
\mu\text{K}$ ($l_c\sim 0.5\,\mu \text{m}$). We will consider therefore only the
interferences between the term $C_{ab}$ and $C_{ba}$ and the expression of
$P_a$ becomes:
\begin{eqnarray}
P_a=\int |\varphi (p)|^2\, (|C_{aa}|^2+|C_{bb}|^2+|C_{ab}|^2+|C_{ba}|^2\nonumber\\
+C_{ab}\cdot C_{ba}^*+C_{ab}^*\cdot C_{ba})\,dp\nonumber\\
\end{eqnarray}
By inserting the equations (\ref{Cij}), (\ref{omega free}) and (\ref{transition
coeff}) in the previous expression and by neglecting the phase terms
proportional to $\frac{\hbar\Delta k^2}{2M}T$ ($\sim 10^{-6}$ for $^{87}Rb$ and
$T=48\,\text{ms}$), we obtain :
\begin{equation}
P_a=P_0+\frac{C}{2} cos((k_{eff}g-\alpha)T^2+\Delta\varphi)
\end{equation}
where $P_0$ is the mean value of $P_a$, $C$ is the contrast and $\Delta
\varphi$ is the phase shift induced by the additional laser lines.
\begin{eqnarray}
P_0=\!\!\left|\sum_{m,m',m''}\!\!\!\!\!\! t^{\frac{\pi}{2}}_m\,t_{m'}^\pi\,t_{m''}^{\frac{\pi}{2}}\,e^{-\theta(m'+2m'')^2}\,e^{i\Delta k(m\,z_A+m'z_C+m''z_F)}\right|^2\nonumber\\
\!\! \!+\left|\sum_{m,m',m''}\!\!\!\!\!\! r_{m}^{\frac{\pi}{2}}\,t_{m'}^\pi\,r_{-m''}^{\frac{\pi}{2}}\,e^{-\theta(m'+2m'')^2}\,e^{i\Delta k(m\,z_A+m'z_B+m''z_D)}\right|^2\nonumber\\
\!\! +\left|\sum_{m,m',m''}\!\!\!\!\!\! t_{m}^{\frac{\pi}{2}}\,r_{m'}^\pi\,r_{-m''}^{\frac{\pi}{2}}\,e^{-\theta(m'+2m'')^2}\,e^{i\Delta k(m\,z_A+m'z_C+m''z_E)}\right|^2\nonumber\\
\!\! +\left|\sum_{m,m',m''}\!\!\!\!\!\! r_{m}^{\frac{\pi}{2}}\,r_{-m'}^\pi\,t_{m''}^{\frac{\pi}{2}}\,e^{-\theta(m'+2m'')^2}\,e^{i\Delta k(m\,z_A+m'z_B+m''z_E)}\right|^2\nonumber
\end{eqnarray}

\begin{eqnarray}\label{C exp phi}
\frac{C}{4}e^{i\Delta\varphi}=&\nonumber\\
\sum_{m,m',m''}& t_{m}^{\frac{\pi}{2}}\,r_{m'}^\pi\,r_{-m''}^{\frac{\pi}{2}}\,e^{-\theta(m'+2m'')^2}\,e^{-i\Delta k(m\,z_A+m'z_C+m''z_E)}\nonumber\\
\times \sum_{m,m',m''}& r_{m}^{\frac{\pi}{2}}\,r_{-m'}^\pi\,t_{m''}^{\frac{\pi}{2}}\,e^{-\theta(m'+2m'')^2}\,e^{i\Delta k(m\,z_A+m'z_B+m''z_E)}\nonumber\\
\end{eqnarray}
with:
\begin{eqnarray}
z_A&=&-z_M\nonumber\\
z_B&=&-z_M+\frac{p_0+\hbar k_{eff}}{M}T+\frac{1}{2}g\,T^2\nonumber\\
z_C&=&-z_M+\frac{p_0}{M}T+\frac{1}{2}g\,T^2\nonumber\\
z_D&=&-z_M+\frac{p_0+\hbar k_{eff}}{M}2T+\frac{1}{2}g\,(2T)^2\nonumber\\
z_E&=&-z_M+\frac{2\,p_0+\hbar k_{eff}}{M}T+\frac{1}{2}g\,(2T)^2\nonumber\\
z_F&=&-z_M+\frac{p_0}{M}2T+\frac{1}{2}g\,(2T)^2\nonumber\\
\theta&=&\frac{\Delta k^2 T^2 k_B T_a}{2M}\nonumber
\end{eqnarray}
In this expression, we constat that the additional laser lines are responsible
for a decrease of the interferometer contrast. For practical case, this
contrast diminution is small ($\sim 3\%$) and does not affect the performance
of the inertial sensor. We find also that the phase of the interferometer is
modified. Without additional laser lines, we obtain the classical result
$(k_{eff}g-\alpha)T^2$. The presence of the additional lines gives an
additional phase shift $\Delta\varphi$ which induces if not corrected an error
on the acceleration measurement. This phase shift depends on the parameters of
the atom interferometer: $\Delta$, $T$, $p_0$, $T_a$ ,$E_n$ , $z_M$. In the
expression ($\ref{C exp phi}$), one can see that the phase shift has a periodic
dependence with the mirror position $z_m$ and the initial velocity $p_0/m$ with
respectively a period of $2\pi/\Delta k $ and $2\pi/(\Delta k T)$.

With the expression (\ref{C exp phi}), it is possible to numerically calculate
the phase shift induced by the additional laser lines. For a $^{87}$Rb atom
interferometer with the parameters $T=48\,\text{ms}$ and
$\Delta/2\pi=0.7\,\text{GHz}$, the phase shift is between $\pm 160\,
\text{mrad}$ depending on the mirror position $z_M$. This phase shift which
corresponds to an acceleration of $\pm 4\times10^{-6}\,\text{m}/\text{s}^{2}$
has to be evaluated precisely in order to reach the state of the art accuracy
of $\sim 10^{-8}\,\text{m}/\text{s}^{2}$ \cite{gravi peter}. And consequently,
the parameters needed in the calculation of the phase shift have to be known
precisely.
\begin{figure}[h]
 \includegraphics[scale=0.4]{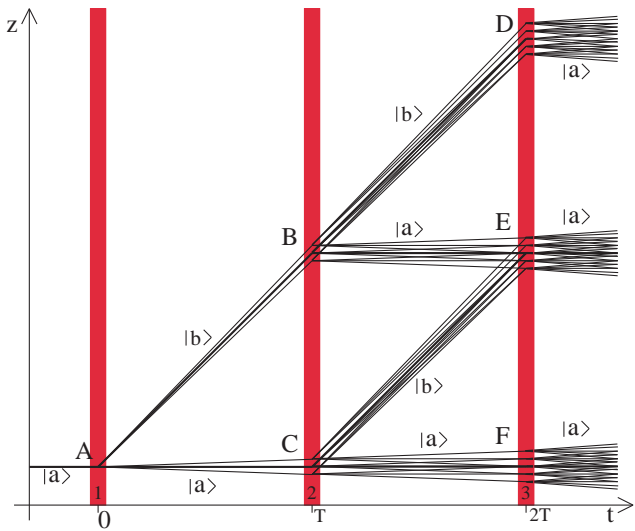}
 \caption{\label{interf} Spatio-temporal diagram of an atomic Mach-Zehnder interferometer using a modulated Raman laser }
\end{figure}

\section{Simplified expression of the phase shift in the limit of low
temperature }

It is possible to simplify the expression of the additional phase
$\Delta\varphi$ by assuming that the spatial separation between parasite paths
$\hbar \Delta k T / m$ is small compared to the atomic coherence length $l_c$ :
\begin{equation}
\theta =\frac{\Delta k^2 T^2 k_B T_a}{2M}\ll 1
\end{equation}
For typical Rubidium atom interferometer parameters ($T=50$ ms, $T_a =2\, \mu
\text{K}$, $G/2\pi=7\, \text{GHz}$), $\theta=0.02$. Within this approximation,
the equation (\ref{C exp phi}) can be simplified and one obtains :
\begin{equation}\label{simplified phase shift}
\Delta\varphi=\varphi_p(z_A)+\varphi_p(z_E)-\varphi_p(z_B)-\varphi_p(z_C)
\end{equation}
where
\begin{equation}
\varphi_p(z)=\arg\left(\sum_m \Omega_m e^{i m \Delta k\, z}\right)
\end{equation}

In the next section, we will validate our model by comparing the calculated
phase shift with experimental measurements.

\section{Comparison with an experiment}

\begin{figure}[h]
 \includegraphics[scale=0.90]{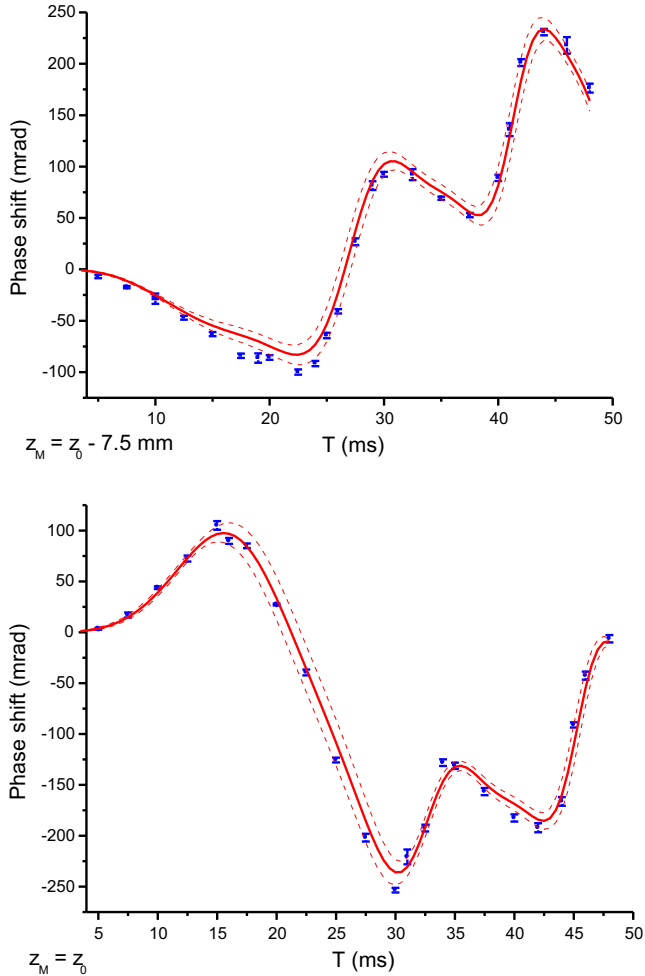}
 \caption{\label{exptheo}Measurement of the phase shift versus T for two different mirror distances and comparison with the theory. The points are the experimental measurements.
  The solid line is the result of the calculation of the phase shift due to the additional laser lines. The only adjustable parameter is the mirror
  distance $z_M$ for the first graph. The dashed lines correspond to the error of the
  calculated phase shift due to experimental uncertainties of the interferometer parameters (see Table \ref{incert phase}).
 }
\end{figure}

Our experimental apparatus consists in a compact atom gravimeter using
$^{87}$Rb. For this atom, the two ground states used for the Raman transition
are $5^2S_{1/2}$ F=1 and F=2 separated in frequency by $6.8\,\text{GHz}$. The
laser used in our atom interferometer is similar to the one described in
\cite{laser onera}. The laser consists in a DFB laser at 1560 nm frequency
doubled in a PPLN crystal. A fiber phase modulator at 1560 nm is used to
generate sidebands at $6.8\,\text{GHz}$ for the Raman transition. The source of
cold atoms is a magneto-optical trap of $^{87}Rb$. After a stage of sub-Doppler
cooling and a Zeeman selection, the atoms are in the state $F=1,\,m_F=0$ at a
temperature of $1.75\,\mu \text{K}$. The atoms interact with a vertical Raman
laser retro-reflected on a mirror placed on a passive vibration isolation
table. The detuning of the Raman laser compared to the excited state
$5^2P_{3/2}$ F'=2 is equal to $-0.7 \,\text{GHz}$. The interferometer sequence
starts 9.5 ms after the beginning of the atoms fall and consists in three
pulses of duration 10, 20 and 10 $\mu\text{s}$ separated by a time $T$ which is
chosen between 5 and 48 ms. During the interferometer a frequency chirp
$\alpha$ is applied to the Raman frequency in order to compensate the Doppler
effect coming from the gravitational acceleration. After the interferometer
sequence, the relative atomic populations in the states F=2 and F=1 are
detected by fluorescence. Interference fringes can be measured by varying the
chirp $\alpha$ applied to the Raman frequency. The gravity acceleration is
obtained by measuring the relative atomic population on each side of the
central fringe ($\alpha_0=k_{\mathrm{eff}} g$) \cite{laser syrte}. In order to
validate our theoretical model described in the previous section, we measured
the gravity for different times T and for two different distances of the mirror
compared to the atoms (see Fig. \ref{exptheo}). Each measurement of gravity
corresponds to an averaging over 1000 atoms drops. Systematic effects on the
gravity measurement depending on T can perturb the comparison theory/experiment
of the phase shift due to additional laser lines. The systematic effects
depending on $k_{\mathrm{eff}}$ (first order light shift, magnetic field) are
canceled by alternating the sign of $k_{\mathrm{eff}}$ \cite{space
application}. The only important systematic effect which does not depend on the
sign of $k_{\mathrm{eff}}$ and has a dependance in T is the two photons light
shift \cite{2 photons, 2 photons syrte}. In the result presented, we take into
account this effect by subtracting to the data the predicted error induced by
the two-photon light shift.

The different parameters used in the calculation of the phase shift induced by
the additional lines have been evaluated. The relative intensities of the laser
lines of the Raman laser $I_n$ have been measured with an optical Fabry-Perot
interferometer at 780 nm. The relative phase of the laser lines are determined
by the fact that the laser is phase modulated : $E_n=\sqrt{I_n}$ for $n>0$ and
$E_n=(-1)^n\sqrt{I_n}$ for $n<0$. By taking into account the two excited states
coupled by the laser with two different Clebsch-Gordan coefficients, we obtain
for the effective Rabi frequencies :
\begin{equation}
\frac{\Omega_n}{\Omega_0}=\frac
{E_{n+1}E_n}{E_1 E_0}\frac{\frac{1}{\Delta_2+nG}+\frac{1/3}{\Delta_1+nG}}{\frac{1}{\Delta_2}+\frac{1/3}{\Delta_1}}
\end{equation}
where $\Delta_{2}$ (resp. $\Delta_1$) is the Raman detuning compared to the
excited state $F'=2$ (resp. 1). The initial velocity of the atoms is deduced
from the gravity and from the delay between the beginning of the atomic fall
and the first Raman pulse. The temperature of the atoms has been measured with
Raman spectroscopy and is equal to $1.75\, \mu  \text{K}$. In our experiment,
we can not evaluate precisely the distance between the atoms and the mirror
($z_M$). This parameter will be adjustable in our calculation in order to
obtain the best fit of our experimental data. For the second set of
measurements, the position of the mirror has been changed by a known quantity
($7.5\,\text{mm}$). Therefore the position of the mirror is not adjusted for
the second set of measurements. For the experimental uncertainties of the
parameters listed in table \ref{incert phase}, the phase shift can be estimated
with an accuracy of 3.1 mrad for the first set of measurements ($z_M=z_0$) and
with an accuracy of 10.3 mrad for the second set of measurements
($z_M=z_0-7.5\,\text{mm}$). We notice that the uncertainty on the phase shift
depends on the mirror distance $z_M$. To obtain the best accuracy on the
acceleration measurement, one has therefore to choose the mirror distance that
minimized the uncertainty of the phase shift. In our case, the optimum distance
is obtained for $z_M \simeq z_0$ (first set of measurement).

\begin{table}[h]
\begin{tabular}{|l|l|l|l|}

\hline
$X$&$\delta X$&$\delta \Delta\varphi$&$\delta \Delta\varphi$\\
&&$(z_M=z_0)$&$(z_M=z_0-7.5 mm)$\\
      \hline
$z_M$&0.3 mm&2.4\,mrad&2.9\,mrad\\
\hline
$p_0/m$&5 mm$\cdot$s$^{-1}$&1.6 mrad&7.7 mrad\\
 \hline
 $T_a$&0.5 $\mu$K&0.96 mrad&0.97 mrad\\
 \hline
 $I_1/I_0$&0.02&0.18 mrad&3.6 mrad\\
 \hline
 $I_{-1}/I_0$&0.02&0.18 mrad&3.7 mrad\\
 \hline
 $I_2/I_0$&0.004&0.05 mrad&1.6 mrad\\
 \hline
 $I_{-2}/I_0$&0.004&0.13 mrad&1.2 mrad\\
 \hline
 $I_3/I_0$&0.0015&0.26 mrad&2.2 mrad\\
 \hline
 $I_{-3}/I_0$&0.0015&0.16 mrad&0.20 mrad\\
 \hline
 \hline
 total& &3.1 mrad&10.3 mrad\\
 &&$(8.3\times 10^{-8}\,\text{m}/\text{s}^{2})$&$(2.8\times 10^{-7}\,\text{m}/\text{s}^{2})$\\
 \hline
\end{tabular}
\caption{\label{incert phase} Experimental uncertainties on the parameters of
the atom interferometer and resulted uncertainties on the estimation of the
phase shift due to additional laser lines ($G/2\pi=6.8\,\text{GHz}$,
$\Delta_2/2\pi=-0.7\,\text{GHz}$, $T=48\,\text{ms}$, $p_0 /
M=9.3\,\text{mm/s}$, $T_a=1.75\,\mu\text{K}$,
$I_{-3,-2,-1,1,2,3}/I_0=0,0.066,0.614,0.636,0.062,0$).}
\end{table}

On Fig. \ref{exptheo}, we compare the measured phase shift on our atom
interferometer with the calculated phase shift induced by the additional laser
lines. We obtain an excellent agreement between theory and experiment. The
differences are compatible with the error bars. We validate our model of phase
shift calculation up to an accuracy of few mrad. For a more precise validation,
one needs a better estimation of the atom interferometer parameters and better
gravity measurements. For our compact gravimeter, the uncertainty of the phase
shift induced by the additional laser lines limits the accuracy of our
acceleration measurement at $8\times10^{-8}\,\text{m}/\text{s}^{2}$. This level
of accuracy is close to the state of the art of
$10^{-8}\,\text{m}/\text{s}^{2}$ and satisfies most of the applications in
gravimetry. In the next part, we will show some methods which reduce the
inaccuracy coming from the use of a modulated Raman.

\section{Methods for reducing the inaccuracy induced by additional laser lines}

If one wants to improve the accuracy limitation induced by the additional laser
lines, a first possibility is to have a better control of the interferometer
parameters. On table \ref{incert phase}, one can see that for our gravimeter,
the inaccuracy is limited by the uncertainty of the mirror position and of the
atoms velocity. The mirror position is not well controlled in our experiment
because the mirror is on an anti-vibration table which can freely move
vertically. If the entire experiment is placed on a vibration isolation table,
the mirror position is fixed and the fluctuation of the mirror distance is only
given by the magneto-optical trap position fluctuations which are in the order
of $20\,\mu \text{m}$ \cite{Ertmer atomic source}. The control of the atoms
velocity can be improved with a Raman velocity selection \cite{Velocity
selection}. A control of the mean velocity and the velocity width under 0.1
mm/s can be easily achieved. With these two improvements of the parameters
control, one obtains for our compact gravimeter ($T=48$ms) a phase shift
uncertainty of 0.45 mrad corresponding to an acceleration accuracy of
$1.2\times10^{-8}\,\text{m}/\text{s}^{2}$.

Another solution to improve the accuracy limitation is to find a configuration
of the interferometer where the phase shift is reduced. On the approximated
expression of the phase shift (\ref{simplified phase shift}), one can see that
the phase shift is equal to zero when $z_A-z_C$ and $z_B-z_E$ are multiple of
the microwave wavelength $2\pi/\Delta k$. As $z_A, z_B, z_C, z_E$ are the
position of the atoms at different points of the interferometer, this condition
is equivalent to say that the distances between the position of the atoms at
the moment of the three Raman pulses is multiple of $2\pi/\Delta k$. The
parameters of the atom interferometer which satisfy this condition are :
\begin{eqnarray}
T=\sqrt{n'\frac{2\pi}{\Delta k\,g}}\\
v_0=\left(n-\frac{n'}{2}\right)\sqrt{\frac{1}{n'}\frac{2\pi\,g}{\Delta k}}
\end{eqnarray}
where $n$, $n'$ are integers.

If this condition is satisfied, the exact phase shift should be minimized. We
perform thus the calculation of the exact phase shift for a time
$T=\sqrt{2\pi/\Delta k\, g}=47.3\,\text{ms}$ and for a velocity
$v_0=0.5\sqrt{2\pi g/\Delta k}=0.232\,\text{m}/\text{s}$ which satisfy the
previous condition. For the experimental parameters described in the previous
section, one obtains a phase shift between $\pm 2.6\,\text{mrad}$ depending on
the mirror position. For the uncertainties of the interferometer parameters
given in table \ref{incert phase} and for the optimum mirror distance, one
obtains an uncertainty on the phase shift estimation of $76\,\mu\text{rad}$
corresponding to an accuracy in acceleration of
$2.0\times10^{-9}\,\text{m}/\text{s}^{2}$. We obtain the same accuracy of
$2.0\times10^{-9}\,\text{m}/\text{s}^{2}$ if we use our atom gravimeter in
micro-gravity \cite{ice nature}. In this case, the atoms are always at the same
position and the phase shift induced by the additional laser lines is also
minimized. With better control of the atom interferometer parameters, the use
of a Raman modulated laser should not be a limitation to reach the extreme
accuracy needed for future space mission \cite{SAGAS, MWXG, STE QUEST}.

\section{Conclusion}

In this article, we have showed that the additional laser lines present in a
Raman laser generated by modulation induce in an atom interferometer a
supplementary phase shift. We have presented a model which allows a precise
calculation of this phase shift. An excellent agreement between experimental
measurements and calculation has validated our model. For our compact
gravimeter, the uncertainty of the parameters needed for the phase shift
calculation limits the accuracy at the level of
$8\times10^{-8}\,\text{m}/\text{s}^{2}$. However, with a better control of the
interferometer parameters or with particular configurations, the inaccuracy is
reduced below $10^{-8}\,\text{m}/\text{s}^{2}$ and does not prevent to reach
the state of the art in gravity accuracy.

The calculation presented here can be easily extrapolated to other inertial
sensors like gyroscopes or gradiometers. In conclusion, the use of a modulated
Raman laser allows to have a laser system more compact and robust and does not
prevent to reach an accuracy at a level of $10^{-8}\,\text{m}/\text{s}^{2}$.

This work was supported by DGA and CNES.


\begin{thebibliography}{}

\bibitem{light pulse inter} J. M. Hogan, D. M. S. Johnson, M. A. Kasevich,
    Proceedings of the International School of Physics "Enrico Fermi" 168, p.
    411-447 (2009).

\bibitem{gravi peter} A. Peters, K.Y. Chung and S. Chu, Metrologia \textbf{38},
    25-61
    (2001).

\bibitem{gravi syrte} A. Louchet-Chauvet, T. Farah, Q. Bodart,
    A. Clairon, A. Landragin, S. Merlet and F. Pereira Dos
    Santos, New Journal of Physics \textbf{13}, 065025 (2011).

\bibitem{gradio} J. M. McGuirk, G. T. Foster, J. B. Fixler, M. J. Snadden, and
    M. A. Kasevich, Phys. Rev. A \textbf{65}, 033608 (2002).

\bibitem{gyro} T. L. Gustavson, P. Bouyer, and M. A. Kasevich, Phys. Rev. Lett.
    78, 2046-2049 (1997);D. S. Durfee, Y. K. Shaham, and M. A. Kasevich, Phys. Rev.
    Lett. \textbf{97}, 240801 (2006).

\bibitem{gyro syrte} A. Gauguet, B. Canuel, T. L\'{e}v\`{e}que, W. Chaibi, and
    A. Landragin, Phys. Rev. A\textbf{ 80}, 063604 (2009).


\bibitem{ice nature} R. Geiger, V. M\'{e}noret, G. Stern, N.
    Zahzam , P. Cheinet , B. Battelier, A. Villing , F. Moron, M.
    Lours, Y. Bidel, A. Bresson, A. Landragin and P. Bouyer, Nature
    Commun. \textbf{2}, 474 (2011).

\bibitem{gradio camion} X. Wu, Gravity Gradient Survey with a Mobile Atom
    Interferometer. Ph.D. thesis, Stanford University, http://atom.stanford.edu/WuThesis.pdf (200 ).

\bibitem{navigation} C. Jeleki, Navigation \textbf{52}, 1-14 (2005).

\bibitem{space} Special Issue: Quantum Mechanics for Space Application: From
    Quantum Optics to Atom Optics and General Relativity, Appl. Phys. B \textbf{84}
    (2006).

\bibitem{SAGAS} P. Wolf et al., Experimental Astronomy \textbf{23}, 651-687
    (2009).

\bibitem{MWXG} W. Ertmer et al., Matter Wave Explorer of Gravity (MWXG), Exp.
    Astron. 23 , 611–649 (2009).

\bibitem{STE QUEST} STE-QUEST Space-Time Explorer and QUantum Equivalence
    Principle Space Test, http://sci.esa.int/ste-quest.

\bibitem{etna} S. Branca, D. Carbone, and F. Greco, Geophysical Research
    Letters \textbf{30}, 2077 (2003).

\bibitem{subsurf det} D. K. Butler, Geophysics \textbf{49}, 1084-1096 (1984).

\bibitem{stimulated raman transition} M. Kasevich and S. Chu, Phys. Rev. Lett.
    \textbf{67}, 181-184 (1991).

\bibitem{laser syrte} P. Cheinet, F. Pereira Dos Santos , T. Petelski, J. Le
    Gou\"{e}t, J. Kim, K.T. Therkildsen, A. Clairon and A. Landragin, Appl. Phys. B
   \textbf{84}, 643-646 (2006).

\bibitem{limit sensitivity grav} J. Le Gou\"{e}t, T.E. Mehlst\"{a}ubler, J.
    Kim, S. Merlet, A. Clairon, A. Landragin and F. Pereira Dos Santos, Appl. Phys. B
    \textbf{92}, 133-144 (2008).

\bibitem{laser onera} O. Carraz, F. Lienhart, R. Charri\`{e}re, M. Cadoret, N.
    Zahzam, Y. Bidel and A. Bresson, Appl. Phys. B \textbf{97}, 405-411 (2009).

\bibitem{space application} A. Landragin, and F. Pereira Dos Santos, in Atom
    Optics and Space Physics, Proceedings of the Enrico Fermi International School of
    Physics "Enrico Fermi," Course CLXVIII, Varenna, 2007, edited by E.
    Arimondo, W. Ertmer, E. M. Rasel, and W. P. Schleich (IOS press) p. 337-350
    (2009).

\bibitem{2 photons} P. Clad\'{e}, E. de Mirandes, M. Cadoret,
    S. Guellati-Kh\'{e}lifa, C. Schwob, F. Nez, L. Julien, and
    F. Biraben, Phys. Rev. A \textbf{74}, 052109 (2006).

\bibitem{2 photons syrte} A. Gauguet, T. E. Mehlst\"{a}ubler, T.
    L\'{e}v\`{e}que, J. Le Gou\"{e}t,
    W. Chaibi, B. Canuel, A. Clairon, F. Pereira Dos Santos, and A. Landragin,
    Phys. Rev. A \textbf{78}, 043615 (2008).

\bibitem{Ertmer atomic source} T. M\"{u}ller, T. Wendrich, M. Gilowski, C.
    Jentsch, E. M. Rasel, and W. Ertmer, Phys. Rev. A \textbf{76}, 063611 (2007).

\bibitem{Velocity selection} M. Kasevich, D. S. Weiss, E. Riis, K. Moler, S.
    Kasapi and S. Chu, Phys. Rev. Lett. \textbf{66}, 2297-2300 (1991).












\end{thebibliography}
\end{document}